\documentclass[aps,prb,twocolumn,superscriptaddress]{revtex4-1}
\pdfoutput=1

\usepackage{graphicx,xcolor}
\usepackage{hyperref}
\usepackage{amssymb}
\usepackage{amsmath}
\usepackage{mathrsfs}
\hypersetup{
    colorlinks,
    linkcolor={blue!},
    citecolor={blue!},
    urlcolor={blue!}
}

\begin{document}

\title{Subgap resonant quasiparticle transport in normal-superconductor quantum dot devices}

\author{J. Gramich}
\email{joerg.gramich@unibas.ch}
\affiliation{Department of Physics, University of Basel, Klingelbergstrasse 82, CH-4056 Basel, Switzerland}
\author{A. Baumgartner}
\affiliation{Department of Physics, University of Basel, Klingelbergstrasse 82, CH-4056 Basel, Switzerland}
\author{C. Sch\"onenberger}
\affiliation{Department of Physics, University of Basel, Klingelbergstrasse 82, CH-4056 Basel, Switzerland}

\date{\today}

\begin{abstract}
We report thermally activated transport resonances for biases below the superconducting energy gap in a carbon nanotube (CNT) quantum dot (QD) device with a superconducting Pb and a normal metal contact. These resonances are due to the superconductor's finite quasi-particle population at elevated temperatures and can only be observed when the QD life-time broadening is considerably smaller than the gap. This condition is fulfilled in our QD devices with optimized Pd/Pb/In multi-layer contacts, which result in reproducibly large and ``clean'' superconducting transport gaps with a strong conductance suppression for subgap biases. We show that these gaps close monotonically with increasing magnetic field and temperature. The accurate description of the subgap resonances by a simple resonant tunneling model illustrates the ideal characteristics of the reported Pb contacts and gives an alternative access to the tunnel coupling strengths in a QD.
\end{abstract}


\maketitle 

Quantum phenomena in nanostructures with a superconductor (S) and a normal metal contact (N) coupled to low-dimensional electron systems like a quantum dot (QD)\cite{Graeber:2004} have recently gained much attention due to potential applications in quantum technology. Especially prominent are transport phenomena at energies below the superconductor's energy gap, $\Delta$, which typically comprise quasi-particle (QP) tunneling and Andreev processes due to Cooper pair transport. These processes result in a large variety of subgap features, for example Majorana Fermions,\cite{Mourik:2012} which might be used for topological quantum computation,\cite{Sarma:2015} Cooper pair splitting\cite{Recher:2001,Hofstetter:2009,Herrmann:2010,Schindele:2012,Das:2012} as a source of entangled electrons, resonant and inelastic Andreev tunneling,\cite{Gramich:2015} or Andreev bound states (ABSs)\cite{Pillet:2010,Dirks:2011,Lee:2014,Schindele:2014} which can be implemented as Andreev qubits.\cite{Bretheau:2013,Janvier:2015} Recent experiments have highlighted the importance to understand in detail the QP excitations in such structures, which, for example, lead to additional subgap features,\cite{Gaass:2014,Ratz:2014} or to a poisoning of the bound state parity lifetime.\cite{Higginbotham:2015}

To identify subgap transport mechanisms, a transport gap much larger than the QD life time, $\Delta \gg \Gamma$, is very beneficial -- a regime which is not easily achieved in S-QD hybrid devices. In addition, a strong suppression of the QP conductance in the subgap regime is required, which is commonly known as a ``clean gap''. While the widely used superconductor Al\cite{Hofstetter:2009,Herrmann:2010,Schindele:2012} has yielded devices with good transport characteristics, long superconducting coherence lengths, $\xi_0$, and more recently also clean gaps,\cite{Chang:2015,Higginbotham:2015,Taupin:2016} it's small gap renders spectroscopic investigations difficult. S-QD devices based on the large-gap superconductor Nb allowed the observation of several fundamental transport processes\cite{Grove-Rasmussen:2009,Gaass:2014,Ratz:2014,Schindele:2014,Gramich:2015} and new effects due to the large critical field.\cite{Fulop:2015} However, Nb has rather short coherence lengths and the devices often exhibit strongly suppressed or ``soft'' gaps\cite{Grove-Rasmussen:2009,Gaass:2014,Fulop:2015} and complex magnetic field characteristics,\cite{Gramich:2015,Fulop:2015} which make normal state control experiments difficult. In contrast, in the superconductor Pb one finds a large bulk coherence length of $\xi_0\sim 90\,$nm, a superconducting gap of $\Delta \sim 1.3\,$meV, and a low critical field of $\sim 80\,$mT.\cite{Poole:2000} In Pb-based devices large transport gaps have already been demonstrated for carbon nanotubes (CNTs) using tunnel barriers\cite{Chen:2009,Bronn:2013, Dirks:2009} and allowed the observation of Cooper pair splitting in graphene.\cite{Borzenets:2015}
Here we present the growth and fabrication of well-defined, reproducible multi-layered Pb-based superconducting contacts to CNTs, which can be easily applied to other materials like graphene or semiconducting nanowires. We demonstrate reproducibly large and clean superconducting transport gaps in CNT QDs with a narrow Pb-based and a normal metal contact. While our fabrication scheme allows for different tunnel coupling strengths of the S contact to the QD due to an implemented Pd contact layer, we focus here solely on QP transport to demonstrate characteristics ideal for spectroscopy experiments. As an example, we report subgap transport resonances that originate from tunneling of thermally excited QPs through a CNT QD. These features were predicted recently\cite{Whan:1996,Pfaller:2013} and reported for experiments in S-QD-S devices,\cite{Gaass:2014, Ratz:2014} whereas the lack of a large superconducting transport gap prohibited their observation in N-QD-S devices.

\begin{figure}[tb]
	\begin{center}
	\includegraphics[width=\columnwidth]{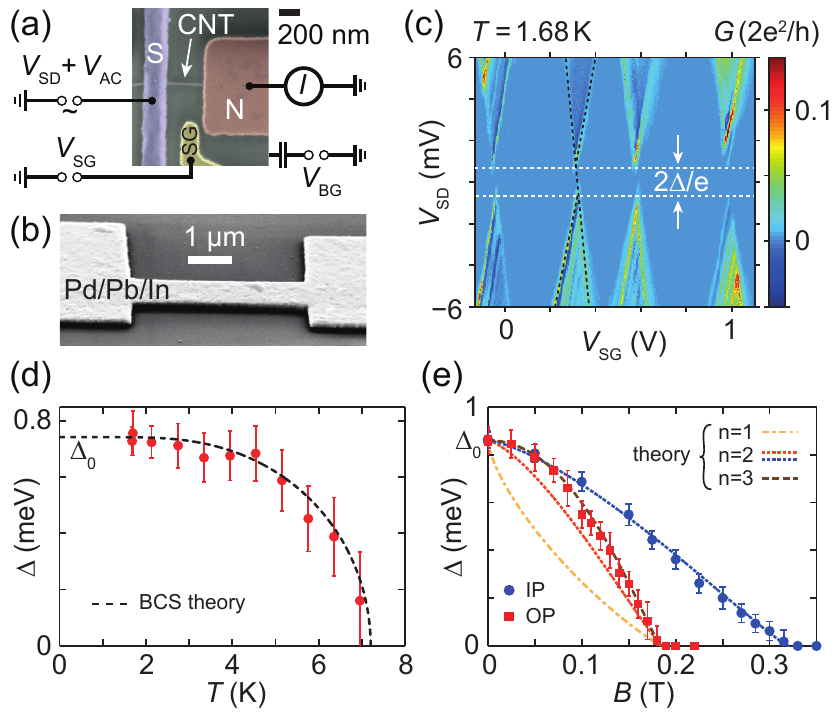}	
\end{center}
	\caption[]{(Color online) (a) False-color SEM image of a typical device with a Pd/Pb/In contact and schematic of the measurement setup. (b) Tilted side-view SEM image of a Pd/Pb/In strip. (c) Differential conductance $G$ of device A as function of $V_\mathrm{SD}$ and $V_\mathrm{SG}$ at $T=1.68\,$K and $V_\mathrm{BG}=-2.987\,$V. The dashed lines mark the onset of QP tunneling and thus the superconducting transport gap $\Delta$. (d) $\Delta$ as function of $T$. The dashed line is the expected dependence from Eq.~\ref{eq:GapEquation}. (e) $\Delta$ of device B as function of the external out of plane (OP, red squares) and in plane (IP, blue dots) magnetic field $B$ at $\sim 30\,$mK base temperature. The dashed lines show the expected dependence in the dirty limit\cite{Skalski:1964} for a pair-breaking parameter $\alpha\propto B^n$. All data in (d) and (e) are extracted from CB spectroscopy and the error bars indicate the individually estimated read-out and statistical errors from 2-4 datasets.}
	\label{fig:DeviceGap}
\end{figure}

Figure~\ref{fig:DeviceGap}(a) shows a false color scanning electron microscope (SEM) image of the N-QD-S device, including a schematic of the measurement setup. CNTs were grown by chemical vapor deposition on a highly p-doped Si/SiO$_2$ substrate used as a backgate (BG). A subsequent surface treatment with radicals from an rf-induced hydrogen plasma\cite{Yang:2010} leads to defect-free, clean CNTs for further processing.\cite{Schindele:2014,Gramich:2015} Using optimized electron beam lithography,\cite{Samm:2014} we fabricate a $\sim 200\,$nm wide S contact and an N contact at a distance of $\sim 300 \,$nm on a CNT, and a single sidegate (SG). We use $50\,$nm of e-beam evaporated Pd for the N contact, SG, and for the outer leads and bonding pads of the narrow S contact. A direct, not optimized evaporation of Pb at room temperature (RT) typically results in a strong island growth, where oxidation between the grain boundaries can result in highly resistive normal conducting Pb strips. Here we deposit an optimized Pd/Pb/In ($4.5-6$/$110$/$20\,$nm) multi-layer \textit{in-situ} as the last fabrication step using electron beam evaporation at a base pressure $<10^{-7}\,$mbar with a Pb deposition rate of $\sim 1.5\,$\AA /s and a sample stage temperature of $\sim 173\,$K. This favors a more uniform Pb growth and reduces Pb surface diffusion. In contrast to the tunnel barriers implemented in Refs.~\citenum{Dirks:2009,Chen:2009,Bronn:2013}, we use a Pd wetting layer to the CNT which allows for some tunability of the S contact coupling strengths and for a smooth and homogeneous Pb growth, see Fig.~\ref{fig:DeviceGap}(b). We employ the superconductor In\cite{Poole:2000} as a capping layer for oxidation protection, which forms a dense and self-limited native oxide layer.\cite{Eldridge:1972} On test strips of the same dimensions as in the CNT devices we determine a critical temperature of $T_\mathrm{c}\approx 7.2-7.4\,$K and a critical out-of-plane (OP) magnetic field of $B^\mathrm{OP}_\mathrm{c} \approx 150-200\,$mT. S-CNT-N devices fabricated in this manner have RT resistances of $\sim 12\,\mathrm{k\Omega} -1\,\mathrm{M\Omega}$, so that different tunnel coupling strengths of the S contacts are feasible. The device characteristics are stable on the timescale of a day under ambient conditions, but the S contacts are damaged during rapid temperature cycling in the cryogenic measurement setup. Here, we focus mainly on experiments performed on device A with a $6\,$nm Pd wetting layer and a RT resistance of $\sim 30\,\mathrm{k\Omega}$. Most measurements employed standard lock-in techniques on a device mounted in a variable temperature insert, allowing experiments at temperatures of $1.5 - 300\,$K. The sample temperature $T$ is determined independently by a LakeShore Cernox resistance thermometer coupled to the device by a copper bridge.

In Fig.~\ref{fig:DeviceGap}(c) the differential conductance $G=\mathrm{d}I/\mathrm{d}V_\mathrm{SD}$ of device A is plotted as a function of the bias $V_\mathrm{SD}$ applied to S and of the sidegate voltage $V_\mathrm{SG}$, at $T=1.68\,$K and the backgate voltage $V_\mathrm{BG}=-2.987\,$V. We observe regular Coulomb blockade (CB) diamonds that are separated due to a well-defined superconducting transport gap, where transport is suppressed for $|V_{\rm SD}|<\Delta_0/e$,\cite{Dirks:2009,Gramich:2015} with $\Delta_0\approx 0.74\,$meV at the lowest sample temperature. We reproducibly find large values of $\Delta_0\approx 0.6-1\,$meV for all 12 measured devices with Pd interlayer thicknesses between $4.5-6\,$nm. Since our devices indicate a reduced $\Delta_0$ with increasing Pd thickness we ascribe the gap reduction from the bulk Pb value ($1.3\,$meV\cite{Poole:2000}) to the proximity effect in the Pd interlayer.\cite{Kim:2012,Garcia:2013} Similarly to Refs.~\citenum{Chang:2015,Taupin:2016} for epitaxial Al-semiconductor nanowires, we find for weakly tunnel-coupled devices a strong ($\sim 100$ times) suppression of the subgap conductance at $T\sim 100\,$mK compared to the normal state ($B>B_\mathrm{c}$) or the above-gap conductance in traces along a CB resonance $\mu_\mathrm{QD}=\mu_\mathrm{N}$ (not shown), for which the QD's electrochemical potential $\mu_\mathrm{QD}$ is aligned with the one of the N contact. This corresponds to a clean and hard superconducting transport gap. $\Delta_0$ seems not to depend on the RT device resistance, nor on the low-temperature tunnel coupling strength, which suggests that $\Delta_0$ is the gap in the metallic Pd-Pb layer. The regular, 2-fold spin-degenerate periodic structure of the CB diamonds in Fig.~\ref{fig:DeviceGap}(c) indicates a clean, defect-free CNT QD, for which a rich substructure of excited states can be resolved due to the sharp QP peaks in the Pb density of states (DOS).\cite{Grove-Rasmussen:2009,Dirks:2009,Gaass:2014} We extract a charging energy of $E_\mathrm{C}\sim 8.7\,$meV and a spacing of the lowest excited states of $\delta E \sim 1.6\,$meV. From CB spectroscopy in the normal state at $T=1.68\,$K, $V_\mathrm{SD}=0$ and $B_\mathrm{OP}=0.6\,{\rm T} >B_\mathrm{c}$, we determine a typical CB resonance width $\sim 0.5\,$meV for device A. Fits with a Breit-Wigner (BW) line shape due to life-time broadening\cite{Gramich:2015} agree well with the data, in spite of the relatively large temperature, with typical tunnel couplings $\Gamma_1 \sim 1-10\,\mathrm{\mu eV}$ and  $\Gamma_2 \sim 500\,\mathrm{\mu eV}$. Since we do not observe Andreev bound states,\cite{Pillet:2010,Dirks:2011,Schindele:2014,Lee:2014} we tentatively ascribe the smaller coupling to S, i.e. $\Gamma_\mathrm{S}=\Gamma_1$. This places device A in the regime $\Gamma_\mathrm{S}\ll \Delta_0 < \delta E \ll E_\mathrm{C}$, in which transport is dominated by Coulomb repulsion and quasi-particle tunneling,\cite{Yeyati:1997,Gaass:2014} while Andreev tunneling\cite{Gramich:2015} is strongly suppressed.

To demonstrate the relevant characteristics of our superconducting Pb contacts, we plot the temperature- and magnetic field dependence of $\Delta$ extracted from individual CB measurements in Fig.~\ref{fig:DeviceGap}(d) and (e). The temperature dependence of device A agrees well (dashed line) with the energy gap obtained from an approximation of the Bardeen, Cooper and Shrieffer (BCS) self-consistency equation\cite{Thouless:1960,DouglassJr:1964}
\begin{equation}
\frac{\Delta(T)}{\Delta_0}=\tanh\left(\frac{T_\mathrm{c}}{T} \frac{\Delta(T)}{\Delta_0}\right),
\label{eq:GapEquation}
\end{equation}
using $\Delta_0=0.74\,$meV and $T_\mathrm{c}=7.2\,$K. This BCS dependence of $\Delta(T)$ is expected to be also approximately valid for the superconductor Pb with a strong electron-phonon coupling.\cite{Gasparovic:1966} The $B$-dependence of $\Delta$ for a similar device B is plotted in Fig.~\ref{fig:DeviceGap}(e), which was measured in a dilution refrigerator at a base temperature of $35\,$mK. At zero field we find $\Delta_0=0.86\,$meV for this device. The field is either applied in-plane (IP) with an in-plane angle of $\sim 15^\circ$ to the Pb strip long axis, or out of plane (OP), i.e. perpendicular to the Pb film. The visible transport gap $\Delta(B)$ is reduced monotonically with increasing $B$ for both cases and vanishes at $B^\mathrm{OP} \sim 180\,$mT and $B^\mathrm{IP}\sim 320\,$mT for the OP and IP configuration, respectively, in reasonable agreement with the critical magnetic fields determined in the resistance measurements on metallic Pb reference strips. From the Pb-layer resistivity $\rho(7.5\,\mathrm{K})\approx 3.9\cdot 10^{-8}\,\rm\Omega m$ determined on reference strips, we estimate a mean free path of $l\approx 50\,$nm, which is comparable to the coherence length $\xi(l)\approx 54\,$nm, but smaller than the penetration depth $\lambda(l)\approx 67\,$nm.\cite{Footnote:1} Because $\lambda > \xi$, the thin Pb films are expected to be type II superconductors. The experimental B-field dependence of the transport gap $\Delta(B)$ is well described by the theory of Ref.~\citenum{Skalski:1964} in the dirty limit $l\ll\xi$ (dashed lines), with a pair-breaking parameter $\alpha\propto B^n$ and exponents $n$ as indicated in Fig.~\ref{fig:DeviceGap}(e).\cite{Footnote:2} Surprisingly, while the expected $n=2$ dependence for IP fields in thin films\cite{Tinkham:2004} agrees well with the data, for the OP field we obtain $n=3$ as best exponent, though $n=1$ is expected in the vortex phase. Here, vortex pinning at Pb island boundaries, the exact local Pb growth configuration and the proximitized Pd layer may play a significant role. Nevertheless, the ideal temperature dependence of $\Delta$ and its monotonic reduction with magnetic field demonstrate that sub-micron Pb contact strips are ideal for transport experiments.
 
As an example for transport spectroscopy in a Pb-based QD system, we now study in some detail the thermally activated QP transport in the transport gap of device A, i.e. for $|V_\mathrm{SD}|<\Delta/e$. If the temperature of a superconductor becomes comparable to the size of the superconducting gap, $kT \sim \Delta$, QPs are excited thermally across the gap with an occupation probability given by the Fermi distribution in S. These QPs can tunnel through the QD to the normal contact and lead to additional subgap transport features, as proposed in Refs.~\citenum{Whan:1996,Pfaller:2013} and found in experiments on S-QD-S devices for both, the sequential\cite{Gaass:2014} and the cotunneling\cite{Ratz:2014} regime. While similar sequential tunneling resonances due to thermally excited QPs have been proposed theoretically also for N-QD-S devices,\cite{Pfaller:2013} no such features were reported so far.

 \begin{figure}[tb]
	\begin{center}
	\includegraphics[width=\columnwidth]{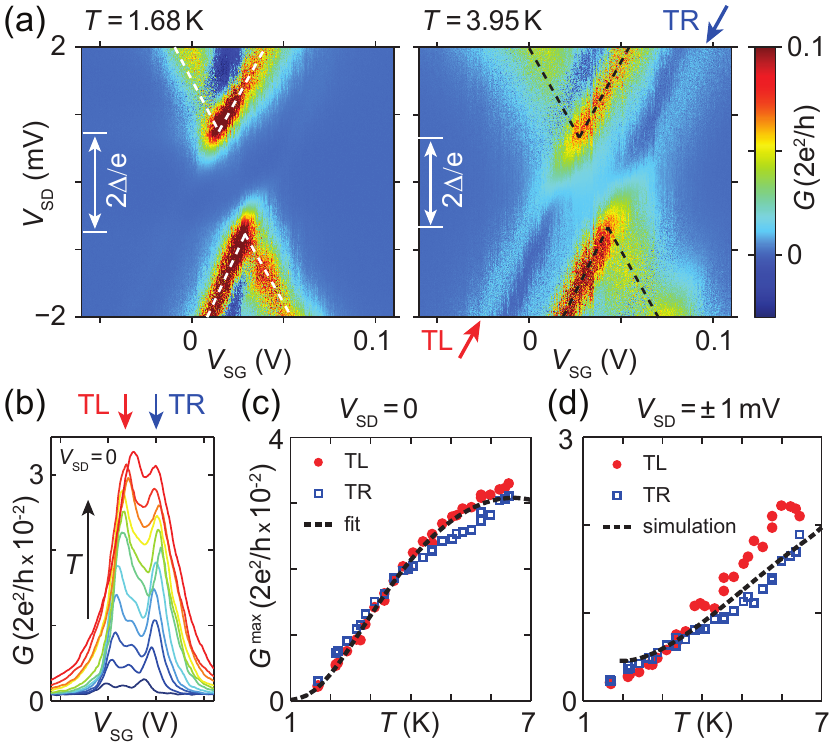}	
\end{center}
	\caption{(Color online) (a) $G$ as function of $V_\mathrm{SD}$ and $V_\mathrm{SG}$ at $T=1.68\,$K (left) and $T=3.95\,$K (right), for $V_\mathrm{BG}=-2.987\,$V. Extra thermal lines (TL/TR, arrows) appear at higher temperatures. (b) Waterfall plot of cross-sections at $V_\mathrm{SD}=0$ in (a) for $T=1.68\,$K (dark blue) to $6.45\,$K (red), extracted from CB spectroscopy with an averaging procedure.\cite{Footnote:4} (c,d) Maximum conductance $G^\mathrm{max}$ of TL (red points) and TR (blue squares) as function of $T$ for (c) $V_\mathrm{SD}=0$ and (d) $V_\mathrm{SD}=\pm 1\,$mV. The dashed line in (c) represents a best fit with Eq.~\ref{eq:ModelCurrent} and fit parameters $\Gamma_1=33\,\mathrm{\mu eV}$, $\Gamma_2=490\,\mathrm{\mu eV}$, the line in (d) a model simulation with the same parameters.}
	\label{fig:ExpQPtransport}
\end{figure}

\begin{figure}[tb]
	\begin{center}
	\includegraphics[width=\columnwidth]{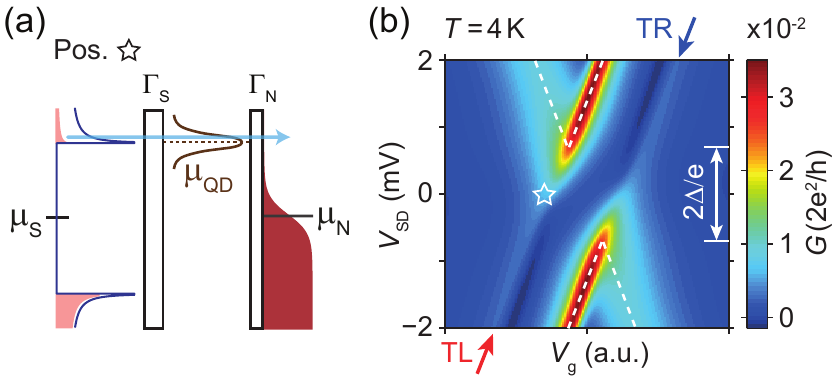}	
\end{center}
	\caption{(Color online) (a) Schematic of thermally activated quasiparticle transport for $kT \sim \Delta$. Thermally excited quasiparticles in S (light red) tunnel through the QD if $\mu_ \mathrm{QD}=\mu_\mathrm{S}+\Delta$ even for $\mu_\mathrm{S}\sim\mu_\mathrm{N}$. (b) Model simulation of $G (V_\mathrm{SD},V_\mathrm{g})$ with Eq.~\ref{eq:ModelCurrent}. Similar to the experiment, extra thermal lines TL and TR (arrows) appear. The star indicates the position of the schematic in (a).}
	\label{fig:TheoryQPtransport}
\end{figure}

Figure \ref{fig:ExpQPtransport}(a) shows a detailed map of $G$ for a CB region of device A as function of $V_\mathrm{SD}$ and the gate voltage $V_\mathrm{SG}$ at $T=1.68\,$K (left) and for an increased temperature of $T=3.95\,$K (right). While we observe only the standard CB diamond edges separated by $\Delta_0$ at the lowest $T$, additional lines (arrows) labeled TL (left) and TR (right) appear for elevated temperatures besides the expected thermal broadening of CB features. At a finite bias $V_\mathrm{SD}$, the conductance maxima of TL and TR are accompanied by regions of negative differential conductance (NDC, dark blue). We study the temperature dependence of these extra lines in cross-sections $G(V_\mathrm{SG})$ at $V_\mathrm{SD}=0$ and $V_\mathrm{SD}=\pm 1\,$mV, shown for $V_\mathrm{SD}=0$ in the waterfall plot of Fig.~\ref{fig:ExpQPtransport}(b). Each curve is an average over a small bias window $\Delta V_\mathrm{SD}=\pm 8\,\mathrm{\mu eV}$ in individual CB spectroscopy measurements using a moving average filter.\cite{Footnote:3} With increasing temperature the amplitude of the features TL and TR increase, while the background is zero due to CB.\cite{Footnote:4} To compare with the model below, we plot in Fig. \ref{fig:ExpQPtransport}(c) and (d) the temperature dependence of the maximum conductance $G^\mathrm{max}$ of TL (red points) and TR (blue squares) for $V_\mathrm{SD}=0$ and $V_\mathrm{SD}=\pm 1\,$mV, respectively, which show a qualitatively different, but distinctive monotonic increase in $G^\mathrm{max}$ with increasing temperature. We ascribe the resonance lines TL and TR to the sequential tunneling of thermally excited QPs in the superconductor, as shown schematically in Fig.~\ref{fig:TheoryQPtransport}(a): at elevated temperatures of $kT \sim \Delta$, the quasi-electron population at $E>+\Delta$ in S (light red) is finite. When the QD's electrochemical potential $\mu_\mathrm{QD}$ is aligned with this population, i.e. $\mu_ \mathrm{QD}=\mu_\mathrm{S}+\Delta$, a current flows even for a bias smaller than $\Delta/e$, resulting in the additional resonance TL tuned by the bias and the gate voltages via the QD resonance condition. Similarly, the resonance TR is due to the condition $\mu_\mathrm{QD}=\mu_\mathrm{S}-\Delta$ for quasi-hole excitations.

We model these QP processes in a simple resonant tunneling picture.\cite{Yeyati:1997} If the bias is applied to S and we neglect superconducting correlations and the charge dynamics on the QD, the current can be approximated as\cite{Yeyati:1997,Gramich:2015}
\begin{align}
I =\frac{e}{h} \int dE &\mathscr{D}_\mathrm{N}(E)\mathscr{D}_\mathrm{S}(E+eV_\mathrm{SD})T_\mathrm{QD}(E) \nonumber \\
&\times [f_\mathrm{N}(E)-f_\mathrm{S}(E+eV_\mathrm{SD})],
\label{eq:ModelCurrent}
\end{align}
with the constant DOS $\mathscr{D}_\mathrm{N}(E)$ in N and a BCS-type DOS in S normalized to the normal state, $\mathscr{D}_\mathrm{S} (E)/\mathscr{D}_\mathrm{N}(E)=|E|/(\sqrt{E^2-\Delta^2})\cdot \Theta{(|E|-\Delta)}$. $f_\mathrm{S/N}(E)=1/(\exp(E/kT)+1)$ are the Fermi functions in the respective contacts and $T_\mathrm{QD}(E)=(\Gamma_1\Gamma_2)/(\Delta E^2 + (\Gamma_1+\Gamma_2)^2/4)$ is the BW transmission function of the QD, with $\Delta E = E-\mu_\mathrm{QD}$, which also accounts for the gating of the QD by the gates (g, voltage $V_\mathrm{g}$) and the contacts. The differential conductance $G=\mathrm{d}I/\mathrm{d}V_\mathrm{SD}$ can then be calculated directly. Figure \ref{fig:TheoryQPtransport}(b) shows the resulting $G$ for $\Delta=0.7\,$meV, $T=4\,$K, $\Gamma_1=10\,\mathrm{\mu eV}$ and $\Gamma_2=500\,\mathrm{\mu eV}$. The model captures the gate voltage and bias dependence of the experiment very well, including the peak-dip structure with negative differential conductance (NDC) next to the TL/TR resonances due to the non-monotonic DOS of S. These results also agree with previous calculations using a microscopic model.\cite{Pfaller:2013} 

To substantiate that the observed subgap features are due to thermal QP tunneling, we now analyze the temperature dependence of TL and TR's resonance amplitudes at zero bias. The corresponding data are plotted in Fig.~\ref{fig:ExpQPtransport}(c). For a zero-width QD resonance $T_\mathrm{QD}(E)=\delta(\Delta E)$ in Eq.~\ref{eq:ModelCurrent}, one finds $G^\mathrm{max} \propto 1/kT\cdot \cosh^{-2}(\Delta/2kT)$ at $V_\mathrm{SD}=0$ for $T\ll T_\mathrm{c}$. Thus, in agreement with a microscopic description,\cite{Gaass:2014} we expect a low-temperature thermally activated characteristics of $G^\mathrm{max}$ as $\sim \cosh^{-2}(\Delta/2kT)$ and a $\sim 1/kT$ decay at larger temperatures $kT \gg \Delta$ well known for sequential tunneling processes. Due to its large superconducting gap, device A is in the regime dominated by the $\cosh^{-2}$ term. To take into account both, the finite width $\Gamma$ of the resonance and the temperature dependence of $\Delta$,\cite{Footnote:5} we fit Eq.~\ref{eq:ModelCurrent} to the data using the BCS temperature dependence of the gap $\Delta(T)$ obtained from Eq.~\ref{eq:GapEquation}. Using $\Delta_0=0.74\,$meV and $T_\mathrm{c}=7.2\,$K determined independently, we obtain the tunnel couplings $\Gamma_1\approx 33\,\mathrm{\mu eV}$ and $\Gamma_2\approx 490\,\mathrm{\mu eV}$ as the only adjustable parameters for the best fit to the data. The fit is shown in Fig.~\ref{fig:ExpQPtransport}(c) as a dashed line, which describes the data very accurately. The extracted coupling parameters agree well with the ones found from independent CB line shape fits in the normal state. This model also reproduces the finite-bias data: inserting the tunnel couplings obtained from the zero bias fit into Eq.~\ref{eq:ModelCurrent}, we obtain the $V_\mathrm{SD}=\pm 1\,$mV amplitudes in a model simulation without additional fit parameters. The resulting curve is plotted as dashed line in Fig.~\ref{fig:ExpQPtransport}(d) and also agrees well with the experiment. We note that for a given temperature, both, the experiment and the model exhibit only a very weak dependence of $G^\mathrm{max}$ on $V_\mathrm{SD}$ for $|eV_\mathrm{SD}| > kT$ in the direction away from the CB diamond edge, see e.g. Fig.~\ref{fig:ExpQPtransport}(a) and Fig.~\ref{fig:TheoryQPtransport}(b).

In conclusion, we demonstrate the growth and fabrication of an optimized Pd/Pb/In layer as narrow superconducting contact for carbon nanotube quantum dot devices, leading to reproducibly large and clean superconducting transport gaps. We illustrate ideal device characteristics, including a BCS-like temperature-dependence and a monotonic closing of the transport gap in magnetic fields. The large observed gaps allow us to identify subgap transport resonances as thermally activated quasiparticle tunneling. Their concise description by a simple resonant tunneling model corroborates a BCS-type density of states for the multi-layer contacts and provides an alternative possibility to determine the QD coupling strengths to the contacts. The implemented Pd coupling layer allows one to access different transport regimes with large and clean proximity gaps, a major advantage for the study of superconducting quantum dot hybrid structures.

This work was financially supported by the Swiss National Science Foundation (SNF), the Swiss Nanoscience Institute (SNI), the Swiss NCCR QSIT, the ERC project QUEST and the EU FP7 project SE$^2$ND.


\begin{thebibliography}{45}
\bibitem{Footnote:4}
\bibinfo{note}{We ascribe the small central subgap
  conductance peak between TL and TR to the thermally broadenend DOS in the S
  contact, coinciding with $\mu_\mathrm{QD}=\mu_\mathrm{N}$.
  The analysis at $V_\mathrm{SD}=\pm 1\,$mV shows that this finite subgap
  conductance at elevated temperatures has no influence on our analysis.}
\bibitem{Footnote:3}
   \bibinfo{note}{We carefully controlled that no structures are lost or created in the averaging
  procedure.}
\end{thebibliography}

\providecommand{\noopsort}[1]{}\providecommand{\singleletter}[1]{#1}%

\end{document}